\documentclass[11pt]{article}
\pdfoutput=1 
\usepackage{jcappub}
\usepackage{graphicx}
\usepackage{color, graphics}
\usepackage{amsmath,amssymb,amsfonts}
\usepackage{hyperref}

\def\beq{\begin{equation}\begin{aligned}}
\def\eeq{\end{aligned}\end{equation}}

\begin{document}

\title{Establishing the Dark Matter Relic Density in an Era of Particle Decays}

\author[a]{Carlos Maldonado}
\author[b,c]{and James Unwin}
\affiliation[a]{Departamento de Fisica, Universidad de Santiago de Chile, Santiago, Chile.}
\affiliation[b]{Department of Physics,  University of Illinois at Chicago, Chicago, Illinois, 60607, USA}
\affiliation[c]{Simons Center for Geometry and Physics, Stony Brook, New York, 11794, USA}

\abstract{If the early universe is dominated by an energy density which evolves other than radiation-like the normal Hubble-temperature relation $H\propto T^2$ is broken and dark matter relic density calculations in this era can be significantly different. We first highlight that with a population of states $\phi$ sourcing an initial expansion rate of the form $H\propto T^{2+n/2}$, for $n\geq-4$, during the period of appreciable $\phi$ decays the evolution transitions to $H\propto T^4$.  The decays of $\phi$ imply a source of entropy production in the thermal bath which alters the Boltzmann equations and impacts the dark matter relic abundance. We show that the form of the initial expansion rate leaves a lasting imprint on relic densities established while $H\propto T^4$ since the value of the exponent $n$ changes the temperature evolution of the thermal bath. In particular, a dark matter relic density set via freeze-in or non-thermal production is highly sensitive to the temperature dependance of the initial expansion rate.   This work generalises earlier studies which assumed initial expansion rates due to matter or kination domination.}

\maketitle
\notoc

\vspace{-2mm}
\section{Introduction}
\vspace{-2mm}

The dark matter freeze-out paradigm, in particular the WIMP miracle, is prized for its simplicity and predictiveness. However, it is relatively straightforward to arrange for significant deviations in the predictions of freeze-out by changing either the particle physics model, or the cosmological history. For instance, standard freeze-out calculations typically assume that decoupling occurs whilst the energy density of the universe is dominated by radiation in which case the expansion rate of the universe is $H\propto T^2$. However, dark matter freeze-out could occur whilst the universe is dominated by some form of energy other than radiation in which case the usual Hubble-temperature relation $H\propto T^2$ is broken. In particular, one possibility which occurs quite naturally in many Standard Model extensions is the case of an early matter dominated period for which $H\propto T^{3/2}$, or an era of particle decays leading to significant entropy production in the thermal bath in which case $H\propto T^{4}$ \cite{Scherrer:1984fd}. The case of dark matter freeze-out during an early period of matter domination was recently highlighted in \cite{Hamdan:2017psw} and freeze-out whilst $H\propto T^{4}$ was studied in \cite{Chung:1998rq,McDonald:1989jd,Giudice:2000ex,Gelmini:2006pw}. 

More generally the early universe could be dominated by the energy density of a population of states $\phi$ evolving as an arbitrary power of the scale factor $\rho_\phi(t)=\rho_\phi(t_I)a^{4+n}$ for  $t_I$ some initial time. This evolution  can also be expressed in terms of the $\phi$ equation of state $\omega_\phi=(n+1)/3$.  Indeed, observational constraints allow the equation of state for the inflaton to take values in $\omega_\phi\in(-1,-1/3)$ between the end of the de Sitter inflationary period and onset the radiation domination \cite{Ackerman:2010he} potentially allowing for $n\in(-4,-2)$ during this transition period. Moreover, periods of non-standard expansion (with a range of rates) are independently motivated in a variety of contexts such as scalars with periodic  potentials \cite{Gardner:2004in,Choi:1999xn}, brane world cosmologies  \cite{Okada:2004nc,Meehan:2014bya}, and scalar-tensor theories \cite{Dutta:2016htz}. An explicit example which allows general $n$ was constructed in \cite{DEramo:2017gpl}, in which the energy density of the universe is dominated by the contributions from a real scalar field $\phi$ with a potential of the form
\begin{equation*}
V(\phi)= \frac{4-2n}{(4+n)^2t_I^2}~{\rm exp}\bigg[(\phi(t_I)-\phi)\sqrt{n+4}\bigg]~.
\end{equation*}
Values of $n\in(-4,2)$ imply positive potentials, corresponding to $\omega_\phi\in(-1,1)$, however considering also negative potentials extends the range to $n>2$ (which was the focus of \cite{DEramo:2017gpl}).

Such early periods of non-standard cosmology can impact the dark matter evolution and thus it is interesting to explore the implications, while remaining agnostic regarding the form and origin of the evolution of the dominant energy density.
Two recent studies by D'Eramo, Fernandez, \& Profumo \cite{DEramo:2017gpl,DEramo:2017ecx} considered the consequences for dark matter if the relic density is established during a period of fast expansion.  These papers focused on scenarios with $n>0$, in which case the expansion rate $H\propto T^{n/2+2}$ is faster than expected from a radiation dominated universe and the energy density in $\phi$ eventually redshifts to a negligible level. Note that the case $n=2$ corresponds to `kination domination', see e.g.~\cite{Spokoiny:1993kt,Ferreira:1997hj,Salati:2002md,Pallis:2005hm}, in which the energy density of the universe is dominated by the kinetic energy of some scalar field  (the $\dot\phi$ term), for instance the inflaton. Conversely, for $n<0$, in order to recover the successes of standard cosmology, one requires that the state which dominates the energy density eventually decays. Specifically, the universe should be dominated by the Standard Model radiation bath at temperates around 10 MeV and below (until matter-radiation equality) so not to spoil the precision predictions of Big Bang nucleosynthesis (see e.g.~\cite{Sarkar:1995dd}).

Here we study the case of dark matter which freezes out or is produced during a period of particle production due to the decays of some (boson or fermion) state $\phi$ which dominates the energy density and under the assumption that $\rho_\phi\propto a^{n+4}$ thus implying an initial expansion rate of the form $H\propto T^{n/2+2}$.  Notably, during this period of $\phi$ decays, entropy is no longer conserved in the thermal bath and this impacts the dark matter relic abundance calculation.  In particular, our work generalizes the earlier papers of  \cite{McDonald:1989jd,Chung:1998rq,Giudice:2000ex,Gelmini:2006pw} which assume that the initial expansion rate corresponds to an early matter dominated phase. Additionally, this work can also be seen as an extension of dark matter freeze-out or production with a general expansion rate, as studied in \cite{DEramo:2017gpl,DEramo:2017ecx}, to include decays of $\phi$. 
This paper is structured as follows: In Section \ref{sec2} we discuss the formulation of the Boltzmann equation without entropy conservation, generalizing the derivations of \cite{Giudice:2000ex}. Using these results, in Section \ref{sec3} we compare different scenarios for setting the dark matter relic abundance and discuss their dependance on the exponent $n$ of the initial expansion rate. Concluding remarks are given in Section \ref{sec4}.


\section{Boltzmann Equations without Entropy Conservation}
\label{sec2}

We start by deriving expressions for the evolution of the different particle populations in the case that the early universe is dominated by a state $\phi$ resulting in an initial expansion rate of $H\propto T^{n/2+2}$. We show that the expansion rate subsequently transitions to $H\propto T^4$ and derive an expression for the maximum temperature of the Standard Model radiation.
The expressions derived reproduce the results of Giudice, Kolb, \& Riotto \cite{Giudice:2000ex} for matter domination prior to decays ($n=-1$) and Visinelli \cite{Visinelli:2017qga} with kination domination prior to decays ($n=2$).

\subsection{Boltzmann equations}
We start in familiar territory by defining the Hubble parameter
\begin{equation}
H^2=\frac{8\pi}{3M_{\rm Pl}^2}\left(\rho_\phi+\rho_R+\rho_X\right)~,
\label{H}
\end{equation}
where the terms indicate the energy density in $\phi$, Standard Model radiation, and dark matter $X$, respectively. The evolution of quantities can be tracked relative to a dimensionless scale factor $A$ defined as 
\beq 
A\equiv\frac{a}{a_I}  =aT_{\rm RH}~.
\label{dimensionlessA}
\eeq
where $a_I$ is an arbitrary initial reference point which is chosen  to be $a_I=1/T_{\rm RH}$ and  $T_{\rm RH}$ is the reheat temperature following $\phi$ decays.
Emulating the analysis of \cite{Giudice:2000ex}, we rewrite the relevant variables as dimensionless quantities, but where now the energy density of $\phi$ evolves as some arbitrary power $a^{4+n}$
\beq
\Phi &\equiv\ \rho_\phi a^{4+n}T_{\rm RH}^n =\frac{\rho_\phi A^{4+n}}{T_{\rm RH}^4}~,
 \qquad
R \equiv\rho_Ra^4  =\frac{\rho_RA^4}{T_{\rm RH}^4}~, 
\qquad
X \equiv n_Xa^3 =\frac{n_XA^3}{T_{\rm RH}^3}~.
\label{dimensionless}
\eeq
With the assumption that $\rho_X=\langle E_X\rangle n_X$, where $n_X$ is the dark matter number density and $\langle E_X\rangle$ is the expected energy of each dark matter state, one can express the Hubble rate in terms of these dimensionless variables to obtain
 \beq
H&= \frac{T_{\rm RH}^2}{M_{\rm Pl}A^{2+n/2}}
\sqrt{\frac{8\pi}{3}\left(\Phi+RA^n+\frac{\langle E_X\rangle X A^{n+1}}{T_{\rm RH}}\right)}~.
\label{Hvar}
\eeq
This recovers the form of $H$ in \cite{Giudice:2000ex} for $n=-1$.
For an expansion rate of the form of eq.~(\ref{Hvar}), the Boltzmann equations which describe the evolution of the number densities can be expressed in a  manner highly reminiscent to  those studied in \cite{Giudice:2000ex,Gelmini:2006pw} 
\begin{align}
\dot{\rho_\phi}+(4+n)H\rho_\phi  &=  -\Gamma_\phi \rho_\phi \label{eqe1}
\\[5pt]
\dot{\rho_R}+4H\rho_R &=\Gamma_\phi \rho_\phi +2\langle E_X\rangle(n_X^2-n_{\rm eq}^2)\langle\sigma v\rangle \label{eqe2}
\\
\dot{n_X}+3Hn_X &= \frac{b}{m_\phi}\Gamma_\phi \rho_\phi - (n_X ^2 - n_{\rm eq}^2)\langle\sigma v\rangle
\label{eqe3}
\end{align}
where $m_\phi$ and $\Gamma_\phi$ are the $\phi$ mass and decay rate to Standard Model. Dotted variables indicate differentiation with respect to time, the quantity $b$ paramaterises the branching ratio of $\phi$ to dark matter, $\langle\sigma v\rangle$ is the thermally averaged dark matter annihilation cross section, and $n_{\rm eq}$ denotes the equilibrium number density of dark matter which has its usual form. Furthermore, note that the $\phi$ decay rate can be expressed in terms of the reheat temperature after $\phi$ decays
\begin{equation}
\Gamma_\phi=\sqrt{\frac{4\pi^3g_*(T_{\rm RH})}{45}}\frac{T_{\rm RH}^2}{M_{\rm Pl}}~.
\label{gammae}
\end{equation}

We next re-express eqns.~(\ref{eqe1})-(\ref{eqe3}) in terms of the dimensionless units of eq.~(\ref{dimensionless}).
Taking first eq.~(\ref{eqe1}), simple substitution yields
\beq
&\dot{A}\frac{d}{dA}\left(\frac{\Phi}{(Aa_I)^{4+n}T_{\rm RH}^n}\right)+(4+n)H\frac{\Phi}{(Aa_I)^{4+n}T_{\rm RH}^n}=-\Gamma_\phi \frac{\Phi}{(Aa_I)^{4+n}T_{\rm RH}^n}~.
\eeq
After some manipulation,  and using that $H=\frac{\dot a}{a}$, this can be simplified to
\beq
\dot{A}\Phi '=\Gamma_\phi \Phi~,
\label{dotA}
\eeq
where the primed variable indicates differentiation with respect to $A$. Furthermore, using eqns.~(\ref{Hvar}) \& (\ref{gammae}) we can express eq.~(\ref{dotA}) in terms of $\Phi'$ as follows
\beq
\Phi '
&=-\sqrt{\frac{\pi^2g_*(T_{\rm RH})}{30}}\frac{\Phi A^{1+n/2}}{\sqrt{\Phi+RA^n+\frac{\langle E_X\rangle X A^{n+1}}{T_{\rm RH}}}}~.
\label{phie}
\eeq
Analogously, eqns.~(\ref{eqe2}) \& (\ref{eqe3}) can be rewritten in a similar fashion to obtain
\beq
R'=&\sqrt{\frac{\pi^2g_*(T_{\rm RH})}{30}}\frac{\Phi A^{1-n/2}}{\sqrt{\Phi+RA^n+\frac{\langle E_X\rangle X A^{n+1}}{T_{\rm RH}}}} 
+\sqrt{\frac{3}{8\pi}}\frac{2\langle\sigma v\rangle\langle E_X\rangle M_{\rm Pl}A^{n/2-1}(X^2-X_{\rm eq}^2)}{\sqrt{\Phi+RA^n+\frac{\langle E_X\rangle X A^{n+1}}{T_{\rm RH}}}}
\label{rade}
\eeq
and
\begin{equation}
\begin{split}
X'=&\sqrt{\frac{\pi^2g_*(T_{\rm RH})}{30}}\frac{b}{m_\phi}\frac{\Phi T_{\rm RH}A^{-n/2}}{\sqrt{\Phi+RA^n+\frac{\langle E_X\rangle X A^{n+1}}{T_{\rm RH}}}} 
-\sqrt{\frac{3}{8\pi}}\frac{\langle\sigma v\rangle M_{\rm Pl}T_{\rm RH}A^{n/2-2}(X^2-X_{\rm eq}^2)}{\sqrt{\Phi+RA^n+\frac{\langle E_X\rangle X A^{n+1}}{T_{\rm RH}}}}.
\label{dme}
\end{split}
\end{equation}
Thus the dimensionless versions of  eqns.~(\ref{eqe1})-(\ref{eqe3}) are, respectively,   eqns.~(\ref{phie})-(\ref{dme}). 


\subsection{Radiation temperature maxima}

We assume that in the early universe the energy density is dominated by $\phi$ and, moreover, we further suppose that the initial radiation bath is negligible. This implies that the initial energy density of $\phi$ can be written
$\rho_\phi(a_I)=(3/8\pi)H_I^2 M_{\rm Pl}^2$
where $H_I\equiv H(a_I)$ is the initial expansion rate (throughout we will use the subscript $I$ to mean the value of a give quantity  at $a=a_I$). In terms of  dimensionless variables this initial condition is
\begin{equation}
\Phi_I=\frac{3H_I^2 M_{\rm Pl}^2}{8\pi T_{\rm RH}^4}, \hspace{15mm} R_I=X_I=0,  \hspace{15mm}  A_I=1.
\label{IC}
\end{equation}
One instance in which such initial conditions could arise, for instance, is immediately after inflation, as in the case of kination domination \cite{Spokoiny:1993kt,Ferreira:1997hj,Salati:2002md,Pallis:2005hm} corresponding to $n=2$.

In  \cite{Giudice:2000ex} the authors studied the evolution of the temperature of the Standard Model thermal bath, assuming that prior to the decays of $\phi$ the radiation component is negligible. What was observed is that during the early matter dominated era the temperature rises due to the decays of $\phi$ until it hits some maximum temperature $T_{\rm Max}$ after which the expansion rate transitions to $H\propto T^4$. The bath then cools until the temperature $T_{\rm RH}$ at $H\simeq \Gamma_\phi$ after which $\phi$ decays become negligible and the universe becomes radiation dominated with $H\propto T^2$.  We next generalise this analysis to the case that the early universe expansion rate sourced by $\phi$ is an arbitrary power of the temperature as in eq.~(\ref{H}).

The $\phi$ decays transfer energy to the Standard Model bath and one can derive the point $A_{\rm Max}$ at which the maximum temperature of the radiation $T_{\rm Max}$ occurs for the more general expansion rate.
Since the dominant contribution comes from $\phi$ at early time, we can neglect the second term in eq.~(\ref{rade}), and using eq.~(\ref{IC}) we obtain
\begin{equation}
R'=\sqrt{\frac{\pi^2g_*(T_{\rm RH})}{30}}\Phi_I^{1/2}A^{1-n/2}.
\end{equation}
Then integrating we  obtain the following 
\begin{equation}
R=
\left\lbrace
\begin{array}{ll}
\sqrt{\frac{\pi^2g_*(T_{\rm RH})}{30}}\sqrt{\Phi_I}\left(\frac{1}{2-n/2}\right)(A^{2-n/2}-1)
 &~~~ \text{for} \ \ \ n<4 \\[10pt]
\sqrt{\frac{\pi^2g_*(T_{\rm RH})}{30}}\sqrt{\Phi_I}\ln(A) &~~~ \text{for}\ \ \ n=4
\end{array}
\right.~.
\label{Radearly}
\end{equation}
The temperature is a measure of the radiation energy density, and thus we can obtain an expression for the evolution of $T$ as a function of $A$ from the expression 
\beq
\rho_R=\frac{\pi^2g_*(T)}{30}T^4=R\left(\frac{T_{\rm RH}}{A}\right)^4~.
\label{Temperaturee}
\eeq 
Moreover, substituting eq.~(\ref{Radearly}) into eq.~(\ref{Temperaturee}) and rearranging, gives the evolution of the temperature (for $n\neq 4$)
\beq
T&=&\left(\frac{45}{4\pi^3}\frac{g_*(T_{\rm RH})}{g_*^{2}(T)}\right)^{1/8}\left(H_IM_{\rm Pl}T_{\rm RH}^2\right)^{1/4}\left[\frac{A^{-(2+n/2)}-A^{-4}}{2-n/2}\right]^{1/4}~.
\label{TTT}
\eeq

The critical point, with respect to $A$, of the factor in square brackets of eq.~(\ref{TTT}) marks the maximum temperature and the value of the scale factor $A_{\rm Max}$ at which the temperature stops increasing and begins to decrease. For  $|n|<4$ this is given  by
\beq 
A_{\rm Max}=\left(\frac{n+4}{8}\right)^{2/(n-4)}~.
\label{amax}
\eeq
For comparison, recall that $A_I=1$ corresponds to $T=T_{\rm RH}$ and $A>1$ implies $a> a_I$. For $A>A_{\rm Max}$ the $A^{-4}$ piece in eq.~(\ref{TTT}) can be neglected and $T\propto A^{-(2+n/2)}$. Observe that $A_{\rm Max}$ is sensitive to the exponent $n$ of the early universe expansion rate.
The temperature extremum  $T_{\rm Max}$ for $|n|<4$ is found at $A=A_{\rm Max}$ given by
\beq
 T_{\rm Max}=
\left(\frac{45g_*(T_{\rm RH})}{4\pi^3g_*^{2}(T_{\rm Max})}\right)^{1/8}\left(M_{\rm Pl}H_I T_{\rm RH}^2\right)^{1/4}\left(\frac{2}{4-n}\right)^{1/4}
\left[\left(\frac{n+4}{8}\right)^{\frac{4+n}{4-n}}-\left(\frac{n+4}{8}\right)^{\frac{8}{4-n}}\right]^{1/4}.
\label{Tmax}
\eeq
As a reference, taking a few specific choices for $n$, the value of $T_{\rm Max}$ can be approximated as
\beq
T_{\rm Max}\simeq
\left(M_{\rm Pl}H_I T_{\rm RH}^2\right)^{1/4}\times\left\lbrace
\begin{array}{ll}
0.30
& \text{~~for} \ \ \ n=-1 \\[8pt]
0.31
& \text{~~for} \ \ \ n=-2 \\[8pt]
0.33
& \text{~~for} \ \ \ n=-3 \\[8pt]
\end{array}
\right.~,
\label{tempapprox}
\eeq
where we take  $g_*(T_{\rm RH})\approx g_*(T_{\rm Max})\approx100$. For example, taking reasonable values for both $H_I\sim$ eV and $T_{\rm RH}\sim 1$ TeV, then $T_{\rm Max}\sim 3$ TeV for $|n|\sim\mathcal{O}(1)$. 
Furthermore, we can re-express eq.~(\ref{TTT}) in terms of a normalised function $f(A_{\rm Max})=1$ as follows
\beq
T= T_{\rm Max} f(A)
\eeq
for 
\beq
f(A)\equiv \kappa(T)\left[A^{-(2+n/2)}-A^{-4}\right]^{1/4}
\eeq
 with
\beq
\kappa(T)=\left[\frac{g_*(T_{\rm Max})}{g_*(T)}\right]^{1/4}
\left[\left(\frac{4 + n}{8}\right)^{\frac{4 + n}{4 - n}}-\left(\frac{4 + n}{8}\right)^{\frac{8}{4 - n}} 
  \right]^{-1/4}~.
\eeq
For reference, a selection of specific values for $\kappa$ are
\beq
\kappa(T)\approx \left[\frac{g_*(T_{\rm Max})}{g_*(T)}\right]^{1/4}
\left\lbrace
\begin{array}{ll}
\left(\frac{8^8}{3^3\cdot5^5}\right)^{1/20} & \text{~~for} \ \ \ n=-1 \\[8pt]
\left(\frac{4^{4}}{3^3}\right)^{1/12} & \text{~~for} \ \ \ n=-2 \\[8pt]
\left(\frac{16^{6}}{7^7}\right)^{1/28} & \text{~~for} \ \ \ n=-3
\end{array}
\right.~.
\label{tempapprox}
\eeq

	\begin{figure}[t!]
		\centering
		\includegraphics[width=0.65\textwidth]{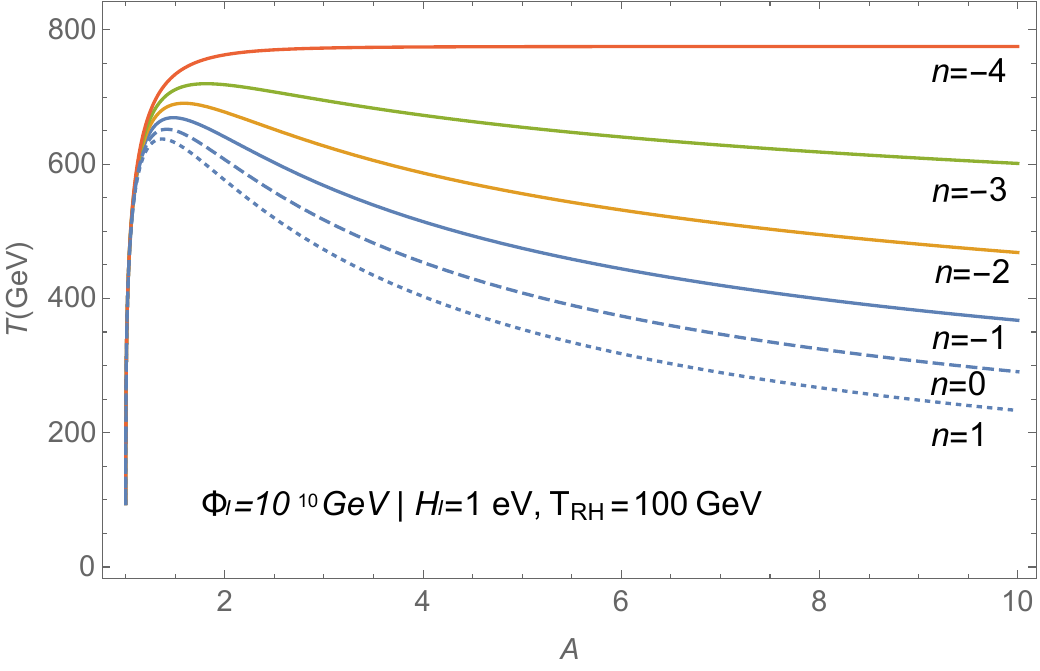}
		\caption{
		The figure shows the bath temperature $T$ as function of $A$ for different values of $n$, assuming that initially the expansion rate is $H\propto T^{n/2+2}$ and that there is negligible energy in the radiation bath or dark matter $R(a_I)=X(a_I)=0$. We fix $\Phi_I=\Phi(a_I)=10^{10}$ GeV, or equivalently (see eq.~(\ref{IC})) this corresponds to, for example, $H_I=1$ eV and $T_{\rm RH}=100$ GeV. The curves follow eq.~(\ref{Tapprox}). Of the cases shown $n=-1$ corresponds to a matter-like $\phi$ (blue, solid), $n=0$ is radiation-like $\phi$ (dashed), and for $n=1$ then $\phi$ redshifts faster than radiation (dotted).  Note that the maximum temperature drops (and occurs earlier) for increasing $n$ following eq.~(\ref{amax}).}
		\label{fig:1}
	\end{figure}

Starting from $a=a_I$ the temperature increases from a negligible value to $T_{\rm Max}$, and then subsequently decreases according to eq.~(\ref{tempapprox}), as illustrated in Figure \ref{fig:1}. The evolution of the bath temperature in  Figure \ref{fig:1} assumes that $\phi$ dominates the energy density, and the evolution will be altered once the energy in radiation becomes comparable to $\phi$, we denote this $A_{\times}$, as we discuss in the next section.
Thus for $A_{\rm Max}<A<A_{\times}$ one can  approximate the temperature evolution as follows
\beq
T\sim \kappa T_{\rm Max} A^{-(2+n/2)/4}~.
\label{Tapprox}
\eeq
Between the time when $T_{\rm Max}$ is reached and the point of radiation domination, at the earlier of $T_{\rm RH}$ or $T_{\times}$, the $\phi$ field energy density scales as $\rho_\phi=\Phi_IT_{\rm RH}^4/A^{4+n}$.  Since the dominant contribution to the Hubble parameter at early times comes from $\rho_\phi$, it follows from eq.~(\ref{Hvar}) and eq.~(\ref{IC}) that for $A_{\rm Max}<A<A_{\times}$ then
\beq
H^2&\simeq&\frac{8\pi}{3M_{\rm Pl}}\frac{\Phi_IT_{\rm RH}^4}{A^{4+n}}=\left(H_IA^{-(4+n)/2}\right)^2~.
\label{Hearly}
\eeq
Moreover, using eq.~(\ref{Tmax}) and eq.~(\ref{Tapprox}), we can express $A$ in terms of the temperature
\begin{equation}
A^{-(4+n)/2}=\left(\frac{4\pi^3(2-n/2)^2g^2_*(T)}{45g_*(T_{\rm RH})}\right)^{1/2}\frac{T^4}{H_IM_{\rm Pl}T_{\rm RH}^2}~.
\label{AinT}
\end{equation}
Substituting eq.~(\ref{AinT}) into (\ref{Hearly}) gives an expression for $H$ for  $A_{\rm Max}<A<A_\times$
\begin{equation}
H=|4-n|\left(\frac{\pi^3g^2_*(T)}{45g_*(T_{\rm RH})}\right)^{1/2}\frac{T^4}{M_{\rm Pl}T_{\rm RH}^2}~.
\label{HinT}
\end{equation}
Thus the point  $A_{\rm Max}$  indicates the $A$ at which the evolution transitions to $H\propto T^4$. Interestingly,  the form of $H$ at this stage is independent of the preceding expansion rate apart from the prefactor, however because the values of $A_{\rm Max}$ and $T_{\rm Max}$ differ the evolution of cosmological abundances still changes for different values of $n$. 

\subsection{Onset of radiation domination}

\begin{figure}[t!]
		\centerline{
		\includegraphics[height=0.35\textwidth]{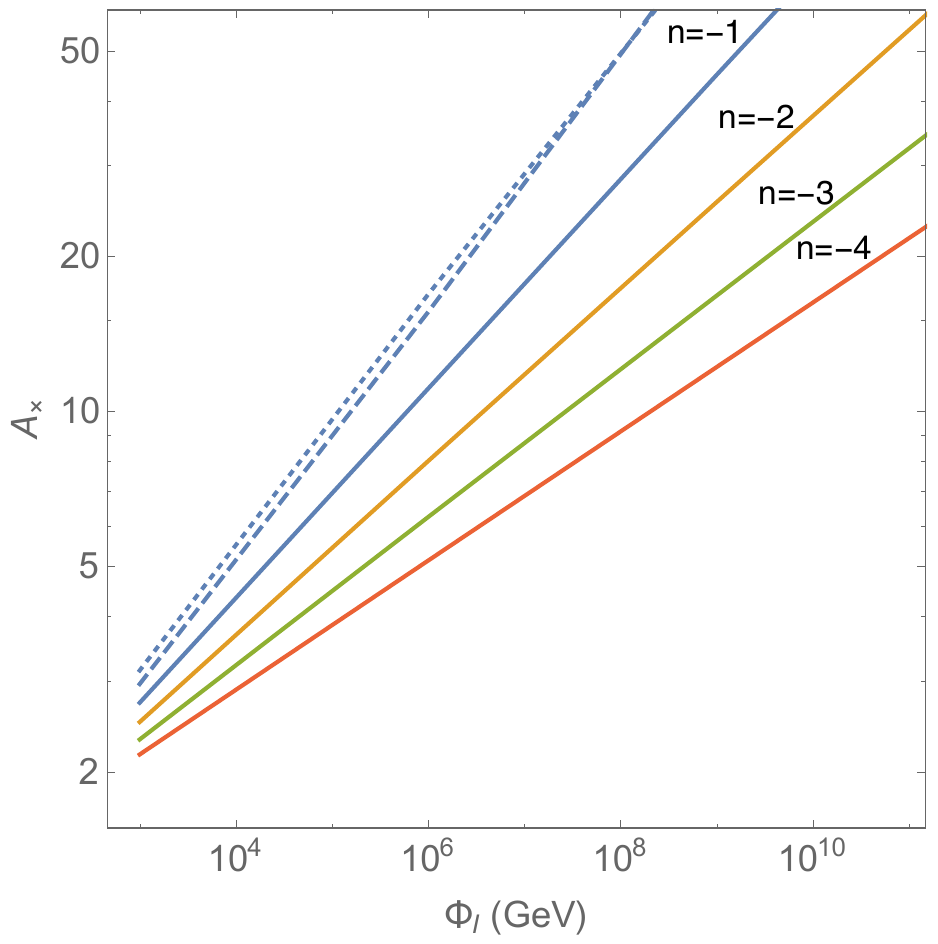}
				\includegraphics[height=0.35\textwidth]{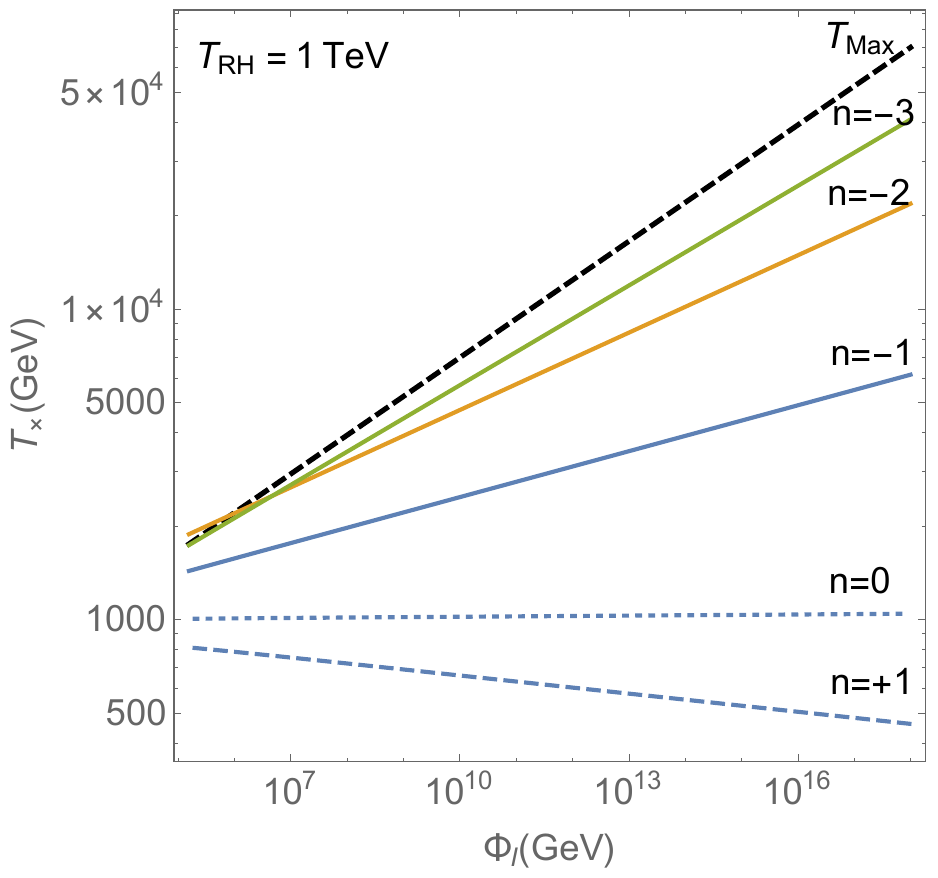}
				\includegraphics[height=0.35\textwidth]{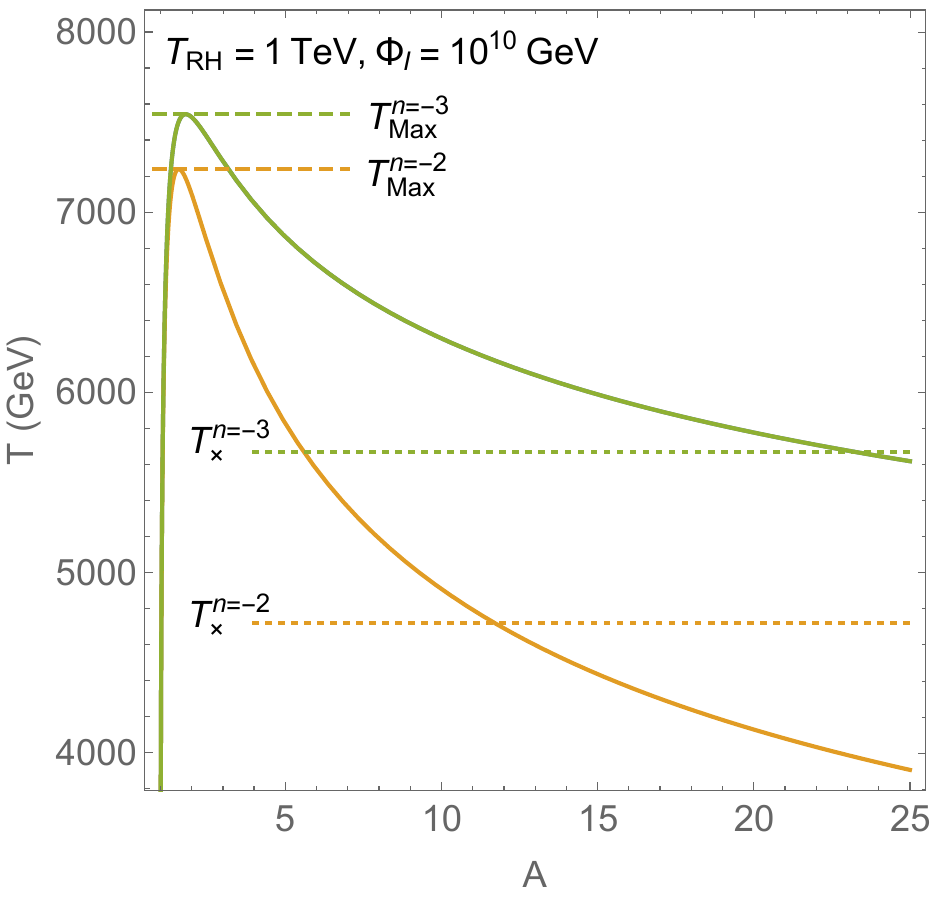}}
		\caption{
		({Left}).~The point $A_{\times}$ at which $R(a_\times)=\Phi(a_\times)$ as $\Phi_I$ is varied and for different $n$, found by solving the coupled differential  eqns.~(\ref{phie}) \& (\ref{rade}) with the initial conditions of eq.~(\ref{IC}).  The line styles match Figure \ref{fig:1}, with $n=0$ is dashed and $n=1$ dotted.  The point $A_{\times}$ signifies the breakdown of eq.~(\ref{phiearly}), which underlies Figure \ref{fig:1}. ({Center}).~For a given value of $T_{\rm RH}$ (the reheating temperature after $\phi$ decay) $A_{\times}$ is associated  to a specific temperature $T_{\times}$ via eq.~(\ref{TTT}). Here we show $T_{\times}$ as a function of $\Phi_I$ for $T_{\rm RH}=1$ TeV. The black dashed curve indicates the maximum temperature $T_{\rm Max}\sim0.3\times (M_{\rm Pl}H_I T_{\rm RH}^2)^{1/4}$. Changes in $T_{\rm RH}$ simply scales the y-axis and the relative orientations of the lines are unchanged. ({Right}).~We illustrate the temperature evolution for two cases  with $n=-2$ and $n=-3$, taking $T_{\rm RH}=1$ TeV and $\Phi_I=10^{10}$ GeV, and we highlight where $T_{\rm Max}$ and $T_{\times}$ occur.
		\label{fig:2} \vspace{-3mm}}
	\end{figure}

Since $\phi$ is decaying eventually radiation will come to dominate the energy density of the universe, indeed this is desirable to match early universe cosmology such as Big Bang nucleosynthesis  observations. The $\phi$ energy density changes as described by  eq.~(\ref{phie}) and at early time (where $X$ and $R$ are negligible) can be expressed as follows  via separation of variables
\beq
\frac{{\rm d \Phi'}}{\sqrt{\Phi}}
&=-
 {\rm d} A 
\sqrt{\frac{\pi^2g_*(T_{\rm RH})}{30}}  A^{1+n/2}~.
\label{phiint}
\eeq
Evaluating this integral from $A_I=1$  we find
\begin{equation}
\Phi=
\left\lbrace
\begin{array}{ll}
{\Phi_I}\cdot {\rm exp}\left[-\sqrt{\frac{\pi^2g_*(T_{\rm RH})}{30}} \frac{1}{2+n/2}(A^{2+n/2}-1)\right]
 &~~~ \text{for} \ \ \ n\not=-4 \\[10pt]
\Phi_I \cdot A^{-{\sqrt{\pi^2g_*(T_{\rm RH})/30}}} &~~~ \text{for}\ \ \ n=-4
\end{array}
\right.~.
\label{phiearly}
\end{equation}
Since the energy density in $\phi$ is  falling quickly, whilst the radiation component  grows gradually, at some point (which we denote $A_{\times}$) the contributions from radiation and $\phi$ become comparable i.e.~$\Phi(A_{\times})\simeq R(A_{\times})$. Shortly after $A_{\times}$ the universe transitions to radiation domination and the expansion rate transitions to $H\propto T^2$.
Importantly, for $A\gtrsim A_{\times}$ then eq.~(\ref{TTT}) (and Figure~\ref{fig:1}) no longer well describe the evolution, since it is not reasonable to neglect $R$ in the derivation.\footnote{This approximation also breaks down if the value of $X$ grows too large, but since the growth of $X$ depends on the small free parameter $b/m_\phi$, we continue to neglect $X$ in deriving $A_{\times}$.}
To find  $A_{\times}$ we numerically solve the coupled differential equations  eqns.~(\ref{phie}) \& (\ref{rade}) with the initial conditions $R(a_I)=X(a_I)=0$. In  Figure \ref{fig:2} (left) we show the values of $A_{\times}$ for different values of $n$ (i.e.~initial expansion rates), and where   $\Phi(a_I)\equiv\Phi_I$ is treated as a free parameter. Fitting to $A_{\times}$ we find the form $A_{\times}=c_n \Phi_I^{m_n}$ where $m_n$ and $c_n$ are constants, for instance, for $n=-1$ then $m_{-1}\approx0.20$ and $c_{-1}\approx 0.68$.

 By inspection of Figure \ref{fig:2} (left) it is seen that for $\Phi_I\gtrsim 10^8$ then eq.~(\ref{phiearly}) is valid up to $A\gtrsim10$, which is the range of Figure \ref{fig:1}. The approximation remains good for lower $\Phi_I$ and higher $A$ for larger values of $n$. Moreover, $A_{\rm Max}$ marks the point of transition from increasing to decreasing bath temperature, thus we should check that $A_{\rm Max}\ll A_{\times}$. Inspecting eq.~(\ref{amax}) we note that for $n\geq-3$ then $A_{\rm Max}\leq2$ and $A_{\rm Max}\ll A_{\times}$ for reasonable values of $\Phi_I$ and $n$.  In particular, for $\Phi_I\gtrsim1$ TeV (typically the range of interest) the bath evolves into the decreasing temperature regime prior to $A_{\times}$. 

For a given $T_{\rm RH}$ one can   translate $A_{\times}$ into the corresponding temperature  $T_{\times}$ at which the radiation and $\phi$ components become comparable, as shown in Figure \ref{fig:2} (center). The approximate value of $T_{\rm Max}\simeq0.3\times (M_{\rm Pl}H_I T_{\rm RH}^2)^{1/4}$ is shown as the  black dashed line and the separation between $T_{\rm Max}$ and $T_{\times}$ gives an indication of the length of the transition from the initial expansion rate to radiation domination.
 Figure \ref{fig:2} (right) illustrates the temperature evolution and the points at which $T_{\rm Max}$ and $T_{\times}$ occur for two specific cases. Once the temperature drops below either  $T_{\times}$ or $T_{\rm RH}$ the system transitions to radiation domination with $H\propto T^2$. The scenario of interest here is the case in which the dark matter relic density is set prior to the onset of radiation domination, as we discuss in the next section.

\vspace{-1mm}
\section{Implications for Dark Matter }
\label{sec3}
\vspace{-1mm}

In the preceding section we studied the behavior of the temperature and the expansion rate for the case of a period in the early universe in which the expansion follows some general power law and while $\phi$ is decaying.  We consider next the implications for dark matter, in particular, how the predicted dark matter relic density depends on the exponent $n$ in the initial expansion rate $H\propto T^{n/2+2}$. We will break the discussion into the following cases:
\begin{itemize}

\item[\bf \S\ref{3.1}:] Freeze-in: Thermal production without chemical equilibrium.

\vspace{-1mm}

\item[\bf \S\ref{3.2}:] Freeze-out during reheating: Thermal production without chemical equilibrium.

\vspace{-1mm}

\item[\bf \S\ref{3.3}:]  Non-thermal production.

\end{itemize}

\subsection{Freeze-in} 
\label{3.1}
First we assume that dark matter is always non-relativistic and does not reach chemical equilibrium   ($X\ll X_{\rm eq}$).  The case of thermal production of non-relativistic dark matter  without reaching chemical equilibrium with the thermal radiation bath is an instance of  dark matter freeze-in  formulated more generally  in \cite{Hall:2009bx} and developed in e.g.~\cite{Elahi:2014fsa,McDonald:2001vt,Chu:2013jja,Yaguna:2011qn,Chu:2011be,Blennow:2013jba}.  
We consider the evolution of $X$ following eq.~(\ref{dme}), for now taking $b=0$, in which case
\begin{equation}
X'=\sqrt{\frac{3}{8\pi}}\langle\sigma v\rangle M_{\rm Pl}T_{\rm RH}\Phi_I^{-1/2}A^{n/2-2}X_{\rm eq}^2~,
\label{X1}
\end{equation}
in terms of the equilibrium distribution given by
\begin{equation}
X_{\rm eq}\equiv a^3n_X^{\rm eq}=\frac{A^3}{T_{\rm RH}^3}g\left(\frac{M_X T}{2\pi}\right)^{3/2}e^{-\frac{M_X}{T}}~,
\label{Xeq}
\end{equation}
where $g$ is the number of internal degrees of freedom of the dark matter state.
Using eq.~(\ref{Tapprox}) and substituting eq.~(\ref{Xeq}) into eq.~(\ref{X1}) we obtain
\beq
X' =\langle\sigma v\rangle \frac{g^2M_X^3\kappa^3T_{\rm Max}^3}{8\pi^3H_IT_{\rm RH}^3}A^{(20+n)/8}e^{-\frac{2M_X A^{(4+n)/8}}{\kappa T_{\rm Max}}}~.
\label{secterm}
\eeq
Expressing the cross section in terms of the s and p-wave pieces $\langle\sigma v\rangle=\alpha_s/M_X^2+ \alpha_pT/M_X^3$, and integrating (neglecting the temperature dependence in $\kappa$, i.e.~$\kappa=\kappa_{\rm RH}\equiv \kappa(T_{\rm RH})$)  gives
\begin{equation}
X_{\infty}=\frac{2^{-(n+28)/(n+4)}}{n+4}g^2\frac{\left(\kappa_{\rm RH} T_{\rm Max}\right)^{(4n+40)/(n+4)}}{\pi^3H_IT_{\rm RH}^3M_X^{24/n+4}}\Gamma\left(\frac{n+28}{n+4}\right)\left(\alpha_s+\frac{n+4}{12}\alpha_p\right)~,
\label{s-pwave}
\end{equation}
 where here $\Gamma$ indicates the gamma function.  Assuming that the dark matter is non-relativistic and does not enter  chemical equilibrium for $b\approx0$ (more precisely provided that the first term of the Boltzmann equation for $X$, eq.~(\ref{dme}), can be safely neglected) then
\begin{equation}
\rho_X(T_{\rm RH})=M_Xn_X(T_{\rm RH})=M_XX_{\infty}\frac{T_{\rm RH}^3}{A_{\rm RH}^3}~.
\label{Xrh}
\end{equation}
Furthermore, it is known that at the point of reheating $H\simeq\Gamma_\phi$ the energy density for radiation is
\begin{equation}
\rho_R(T_{\rm RH})=\frac{\pi^2g_*(T_{\rm RH})}{30}T_{\rm RH}^4
\label{radrh}
\end{equation}
and thus comparing the ratio of energy densities now and at reheating we have
\begin{equation}
\frac{\rho_X(T_{\rm now})}{\rho_R(T_{\rm now})}=\frac{T_{\rm RH}}{T_{\rm now}}\frac{\rho_X(T_{\rm RH})}{\rho_R(T_{\rm RH})}=\frac{M_X}{T_{\rm now}}\frac{30}{A_{\rm RH}^3\pi^2g_*(T_{\rm RH})}X_{\infty}~.
\label{3.7}
\end{equation}
Substituting  eq.~(\ref{s-pwave}) into eq.~(\ref{3.7}) it follows that 
\begin{equation}
\frac{\rho_X(T_{\rm now})}{\rho_R(T_{\rm now})}
= \frac{30\times 2^{-\frac{n+28}{n+4}} M_X}{A_{\rm RH}^3\pi^5g_*(T_{\rm RH})T_{\rm now}(n+4)}\frac{g^2\left(\kappa_{\rm RH} T_{\rm Max}\right)^{\frac{4n+40}{n+4}}}{H_IT_{\rm RH}^3M_X^{\frac{24}{n+4}}}\Gamma\left(\frac{n+28}{n+4}\right)\left(\alpha_s+\frac{n+4}{12}\alpha_p\right)~.
\end{equation}
Using the form of $A_{\rm RH}$ from eq.~(\ref{Tapprox}) we can re-express the dark matter relic abundance as
\begin{equation}
\frac{\Omega_{X}h^2}{\Omega_{R}h^2}
=\frac{30\times 2^{-\frac{n+28}{n+4}}g^2}{\pi^5(n+4)H_Ig_*(T_{\rm RH})T_{\rm now}}\Gamma\left(\frac{n+28}{n+4}\right)
\frac{ (\kappa_{\rm RH}T_{\rm Max})^{4}}{T_{\rm RH}^{3\frac{(n-4)}{(n+4)}}M_X^{\frac{20-n}{n+4}}}\left(\alpha_s+\frac{n+4}{12}\alpha_p\right)~,
\label{relicFI}\end{equation}
in terms of the observed fractional energy densities $\Omega_{R,X}$ for radiation and dark matter. One could further rewrite eq.~(\ref{relicFI}) in terms of $T_{\rm RH}$ by substituting the form of $T_{\rm Max}$ from eq.~(\ref{Tmax}). For $n=-1$ this scenario is studied in `Case A' of  \cite{Giudice:2000ex},  and eq.~(\ref{relicFI}) generalises this to other values of the exponent $n$, reproducing the earlier result for $n=-1$.

	\begin{figure}
		\centerline{
		\includegraphics[height=0.45\textwidth]{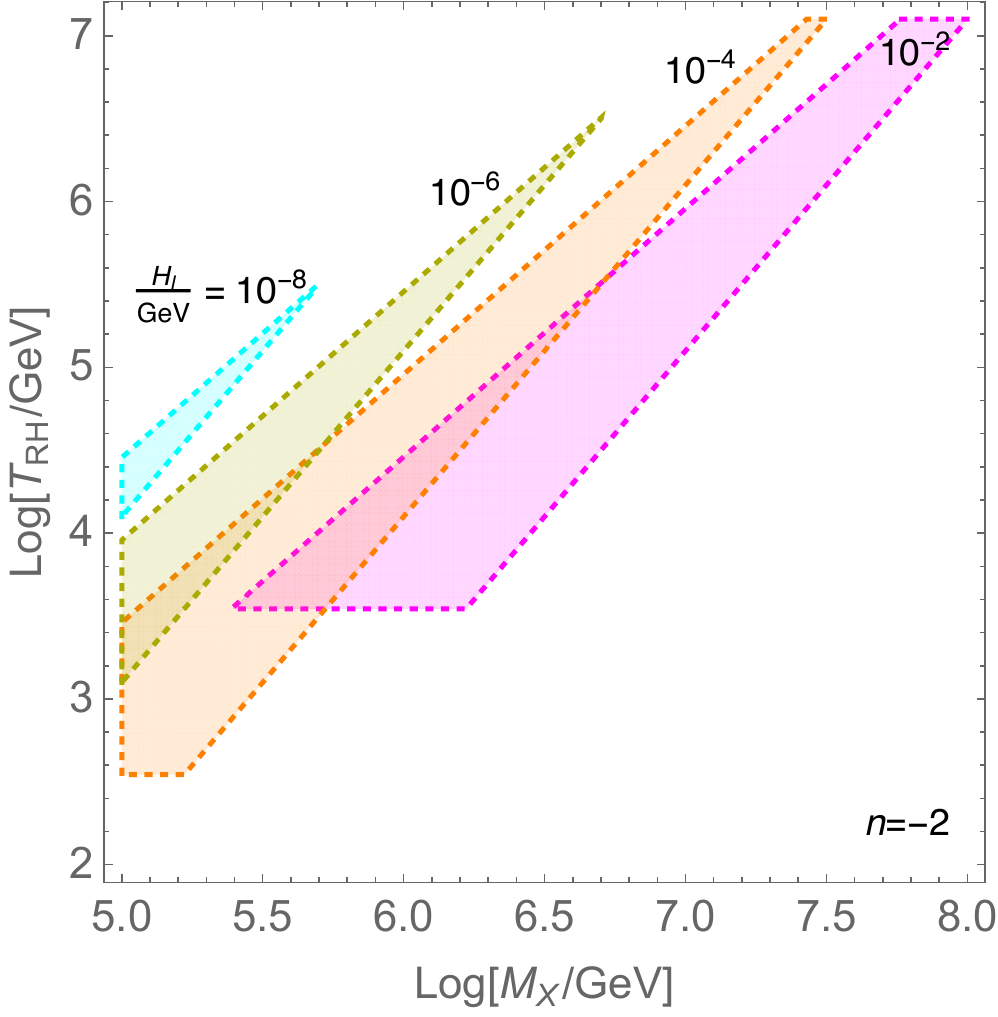}
		\hspace{3mm}
		\includegraphics[height=0.45\textwidth]{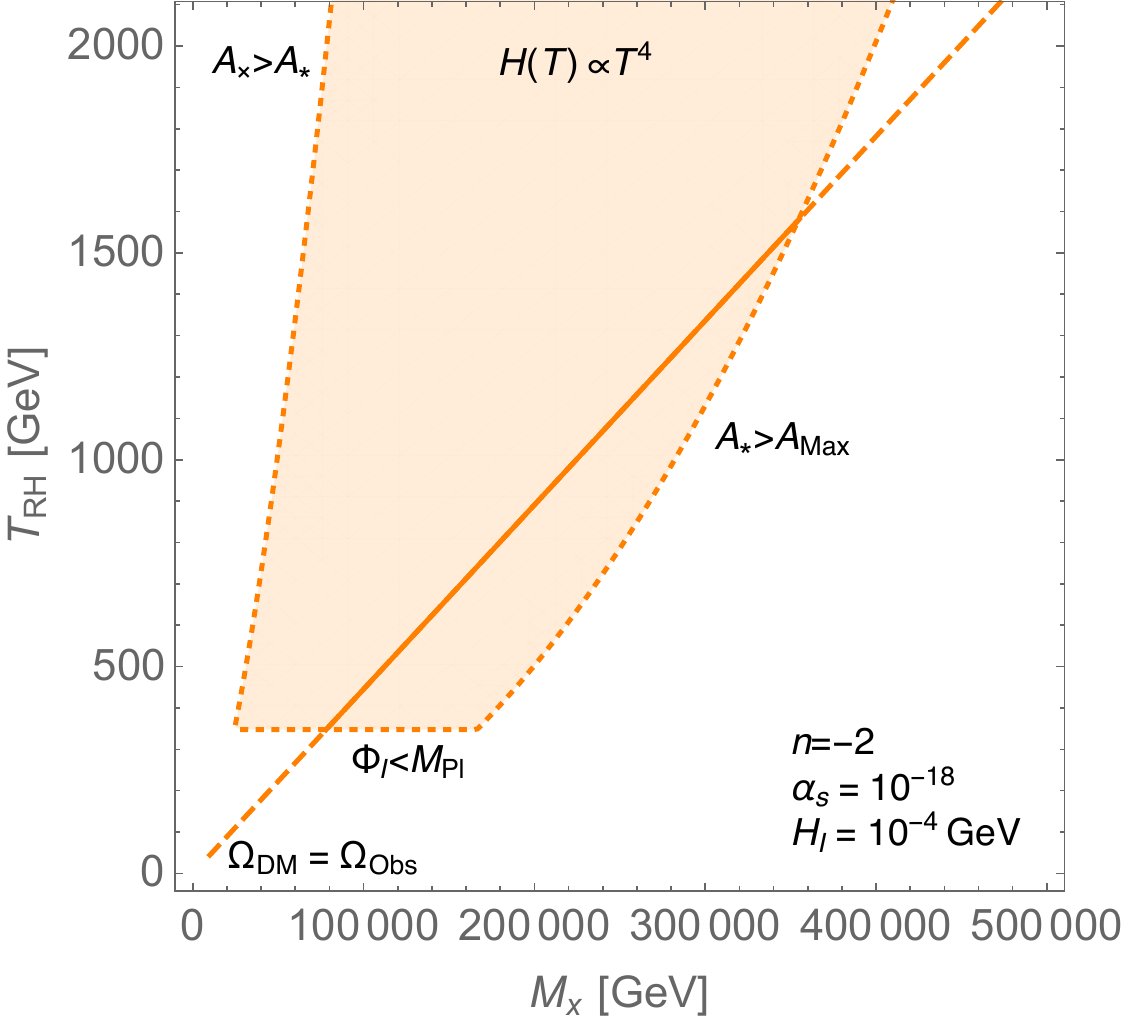}}
			\caption{(Left).~Shaded areas indicate regions of the $T_{\rm RH}$-$M_X$ plane for which $A_{\rm Max}<A_*<A_{\times}$ for $n=-2$ and different values of the initial   expansion rate $H_I$ and we also require  $\Phi_I(H_I,T_{\rm RH})<M_{\rm Pl}$. (Right).~As an example we fix $n=-2$ and $H_I=10^{-4}$ GeV and plot the $T_{\rm RH}$ which gives the observed relic density $\Omega_{\rm Obs}$ by freeze-in as $M_X$ is varied assuming $\alpha_s=10^{-18}$ and $\alpha_p=0$, as described in eq.~(\ref{relicFI}). We highlight the region $A_{\rm Max}<A_*<A_{\times}$ and $\Phi_I<M_{\rm Pl}$ for  $H_I=10^{-4}$ GeV (matching the left panel), outside of this region the relic density curve is unreliable and we indicate this by dashing the line. Note that  $T_{\rm RH}\sim1$ TeV  and $H_I\sim10^{-4}$ GeV corresponds to $\Phi_I\sim10^{17}$ GeV.
		\label{fig:3} \vspace{-3mm}}
	\end{figure}

Additionally, note that the point of peak dark matter production $A_*$ can be found by looking at where the derivative in  eq.~(\ref{secterm}) vanishes. 
For the s-wave case this occurs for\footnote{Note taking $n=-1$ we find that the prefactor is $19/3$ which is slightly different from derived in \cite{Giudice:2000ex} which give the prefactor as $17/2$. We believe the authors use different criteria and approximations.}
\begin{equation}
A_*=\left[\left(\frac{20+n}{4+n}\right)\frac{\kappa T_{\rm Max}}{2 M_X}\right]^{\frac{8}{4+n}}\simeq 
\left(\Phi_I\right)^{\frac{1}{4+n}}   \left[0.3\cdot  \frac{T_{\rm RH}}{2M_X}\left(\frac{20+n}{4+n}\right)\right]^{\frac{8}{4+n}},
\label{aswave}
\end{equation}
where in the final equation we have used eqns.~(\ref{IC}) \& (\ref{tempapprox}). Observe that $A_*$ depends strongly on $T_{\rm RH}/M_X$ but is relatively insensitive to $\Phi_I$. 

For the relic density to be described by eq.~(\ref{relicFI}), i.e.~while $H\propto T^4$, it is required that $A_{\rm Max}<A_*<A_{\times}$. Recall from Figure \ref{fig:2} (left) that $A_{\times}\sim\mathcal{O}(10)$, thus dark matter production  occurs prior to the onset of radiation domination for $A_*<A_{\times}\sim\mathcal{O}(10)$.  Typically it can be arranged that  $A_*\sim\mathcal{O}(10)<A_{\times}$ by choosing an appropriate $\Phi_I$, although requiring that $\Phi_I<M_{\rm Pl}$ introduces an additional restriction.  In Figure~\ref{fig:3} (left) we highlight parameter values for which $A_{\rm Max}<A_*<A_{\times}$ for the case of $n=-2$.  Note that if $\Phi_I$ is relatively large then the dark matter mass must be fairly heavy to reproduce the observed relic density.

In Figure~\ref{fig:3} (right) we show the parameter region  $A_{\rm Max}<A_*<A_{\times}$ along with curve for which the observed dark matter relic density is reproduced ($\Omega_X=\Omega_{\rm Obs})$ for a specific example taking $n=-2$ and $H_I=10^{-4}$ with couplings $\alpha_s=10^{-18}$ and $\alpha_p=0$. We highlight that generally for the relic density curve to align with the region $A_{\rm Max}<A_*<A_{\times}$ diminutive couplings are needed $\alpha\ll1$, however this is actually fortuitous since such feeble coupling strengths are required in freeze-in models in order to ensure that the dark matter remains out of equilibrium \cite{Hall:2009bx} (as assumed for this case). In the next subsection we shall derive the condition on $\alpha$ under which $X< X_{\rm eq}$ at all times.

\subsection{Freeze-out during reheating}
\label{3.2}

Next we consider the case in which the dark matter reaches chemical equilibrium and then freezes out while $H\propto T^4$. The point of freeze-out can be defined implicitly by 
\begin{equation}
n_X^{\rm eq}(T_F)\langle\sigma v\rangle=H(T_F)~.
\label{caseB}
\end{equation}
Using that $n_X^{\rm eq}=X^{\rm eq}A^{-3}T_{\rm RH}^3$ and eq.~(\ref{Xeq}) we can rewrite the lefthand  side of the above equation in terms of $T_F$ and for the righthand side we substitute eq.~(\ref{HinT}) to obtain
\begin{equation}
\frac{g}{\sqrt{8\pi^3}}  \langle\sigma v\rangle (M_X T_F)^{3/2}\exp\left(\frac{-M_X}{T_F}\right)=
|4-n|\left(\frac{\pi^3g^2_*(T_F)}{45g_*(T_{\rm RH})}\right)^{1/2}\frac{T_F^4}{M_{\rm Pl}T_{\rm RH}^2}~.
\end{equation}
Thus the calculation is largely unchanged from earlier studies, differing only in the factor $|4-n|$, and the freeze-out temperature is analogous to as derived in \cite{Giudice:2000ex}, given by
\beq
x_F&=&\ln\left(\frac{3gM_{\rm Pl}T_{\rm RH}^2g_*(T_{\rm RH})^{1/2}}{|4-n|\sqrt{5}\cdot 8\pi^3g_*(T_F)M_X^3}\left(\alpha_sx_F^{5/2}+\alpha_px_F^{3/2}\right)\right)~.
\eeq
Note that, as usual, because of the insensitivity of the logarithm dependences for a large range of reasonable parameter values $x_F\sim \mathcal{O}(10)$. Comparing $T_F\sim M_X/\mathcal{O}(10)$ to $T_{\times}$ in Figure \ref{fig:2} (centre) one finds that $T_\times\sim(\mathcal{O}(100)-\mathcal{O}(1000)$) GeV therefore for $M_X\gtrsim 1$ TeV then typically $T_F > T_\times$. Thus for a large range of parameters dark matter freeze-out can occur well before the transition to radiation domination, while $H$ is described  by eq.~(\ref{HinT}).

The dark matter abundance remains constant after the point of reheating at $H\simeq \Gamma_\phi$. To ascertain the abundance of dark matter at reheating it is necessary to evolve the freeze-out abundance from $T_F$ to $T_{\rm RH}$ as follows
\begin{equation}
\rho_X(T_{\rm RH}) = \left(\frac{a(T_{\rm RH})}{a(T_F)}\right)^{-3} \rho_X(T_F)
 = \left(\frac{g_*(T_{\rm RH})}{g_*(T_F)}\right)^2 \left(\frac{T_{\rm RH}}{T_F}\right)^8 \rho_X(T_F)~,
\end{equation}
where we use that the ratio of FRW scale factors can be replaced by the ratio of dimensionless $A$ factors. 
It follows that the dark matter relic density is given by 
\begin{equation}
\frac{\Omega_Xh^2}{\Omega_Rh^2} =|4-n|\frac{5\sqrt{5}}{4\sqrt{\pi}}\frac{\sqrt{g_*(T_{\rm RH})}}{g_*(T_F)}
\frac{T_{\rm RH}^3}{T_{\rm now}M_XM_{\rm Pl}}
\frac{1}{\alpha_sx_F^{-4}+\alpha_px_F^{-5}/5}~.
\end{equation}
Notably, the abundance is essentially insensitive to the expansion rate prior to reheating in the case that dark matter freeze-out is non-relativistic and in thermal equilibrium.

Whether the relic density is set via freeze-out or freeze-in depends on if the dark matter enters equilibrium. Specifically,  for $X_\infty\lesssim X_{\rm eq}(T_{*})$, where $T_*$ is the temperature of dominant particle production given by eq.~(\ref{TTTTT}),  the dark matter will remain out of equilibrium at all times and the production rate  sets the relic density (the freeze-in scenario). Thus there is a critical value of the coupling $\alpha^{\rm (crit)}$ above which  the inequality $X_\infty\lesssim X_{\rm eq}(T_{*})$ is violated.
From eqns.~(\ref{Tapprox}) \& (\ref{aswave}) we have that $A_*$ corresponds to a temperatures $T_*$ which for s-wave is
\begin{equation}
T_*\simeq M_X\left(\frac{8+2n}{20+n}\right)~.
\label{TTTTT}
\end{equation}
Applying the criteria that for $\alpha=\alpha^{\rm (crit)}$ then $X_\infty= X_{\rm eq}(T_{*})$, it follows that for the case with $\langle\sigma v\rangle\simeq\alpha_s/M_X^2$ the critical coupling $\alpha_s^{\rm (crit)}$ is given by
\beq
\alpha_s^{\rm (crit)}=\frac{2\pi^3 M_X^3\left(20+n\right)^{\frac{3(12-n)}{2(n+4)}}\left(4+n\right)^{\frac{5n-28}{2(n+4)}}|4-n|}{\sqrt{45}g\Gamma\left(\frac{28+n}{4+n}\right)e^{\frac{20+n}{2(n+4)}}M_{\rm Pl}T_{\rm RH}^2}\frac{g_*(T_*)}{g_*(T_{\rm RH})^{1/2}}.
\eeq
Thus for $\alpha_s<\alpha_s^{\rm (crit)}$ the relic density is set by freeze-in, as in Section \ref{3.1}, whereas for  $\alpha_s>\alpha_s^{\rm (crit)}$ freeze-out dynamics determines the dark matter relic density.

\subsection{Non-thermal production}
\label{3.3}

 The scenario of non-thermal production is important for  $b\not\approx0$, in which case a significant (possibly dominant) population of dark matter may be produced directly from $\phi$ decays \cite{Gelmini:2006pw}. The case of non-thermal production without chemical equilibrium is described by eq.~(\ref{dme}) with the second ($b$-independent)  term neglected
\beq
X'=&
\sqrt{\frac{\pi^2g_*(T_{\rm RH})}{30}}\frac{b}{m_\phi}\frac{\Phi T_{\rm RH}A^{-n/2}}{\sqrt{\Phi+RA^n+\frac{\langle E_X\rangle X A^{n+1}}{T_{\rm RH}}}} ~.
\eeq
Integrating (for $n\not=2$) from $A_I$ to $A_{\rm RH}$ and applying the boundary conditions of eq.~(\ref{IC}) we find the total population of dark matter produced due to $\phi$ decays is\footnote{For  $n=2$ then rather $X(T_{\rm RH})\propto \ln(A_{\rm RH})$, this case was studied in \cite{Visinelli:2017qga} and we will not discuss it further.} 
\begin{equation}
X_{\rm RH}\equiv X(T_{\rm RH})\simeq-\frac{2\eta}{n-2}\sqrt{\frac{\pi g_*(T_{\rm RH})}{80}}\frac{H_IM_{\rm Pl}}{T_{\rm RH}}\left(A_{\rm RH}^{1-n/2}-1\right)~,
\label{branratio-nnot2}
\end{equation}
where we define $\eta\equiv b/m_\phi$ which parmaterises the $\phi$-dark matter branching fraction. 

Using the above equation and  eq.~(\ref{Xrh}) \& (\ref{radrh}) it follows that
\beq
\frac{\rho_{X_b}(T_{\rm now})}{\rho_R(T_{\rm now})}
\simeq\frac{15M_XH_IM_{\rm Pl}\eta}{\sqrt{5}(2-n)\pi^{3/2}g_*(\sqrt{T_{\rm RH}}) T_{\rm RH}^4 T_{\rm now}}
\left(A_{\rm RH}^{\frac{2-n}{2}}-1\right)\frac{T_{\rm RH}^3}{A_{\rm RH}^3}~.
\eeq
Further, using. eq.~(\ref{Tapprox}) to replace $A_{\rm RH}$ this can be rewritten to obtain an expression for the dark matter relic abundance
\beq
\frac{\Omega_{X}}{\Omega_R}
&\simeq&\frac{15M_XH_IM_{\rm Pl}\eta}{\sqrt{5}(2-n)g_*(\sqrt{T_{\rm RH}})\pi^{3/2}T_{\rm RH}^3T_{\rm now}}\left(\left[\frac{\kappa T_{\rm Max}}{T_{\rm RH}}\right]^{\frac{4(2-n)}{n+4}}-1\right)\left(\frac{T_{\rm RH}}{\kappa T_{\rm Max}}\right)^{\frac{24}{n+4}}.
\label{nontrd}
\eeq
Similar to previously, the above result generalises expressions in Case 3 of \cite{Gelmini:2006pw} from $n=-1$ to general $n$.

Note that the preceeding calculation assumes the ordering $A_{\rm Max}<A_{\rm RH}<A_{\rm \times}$ where $A_{\rm RH}\equiv A(T_{\rm RH}).$ We can obtain an expression for $A_{\rm RH}$ from eqns.~(\ref{gammae}) \& (\ref{TTTTT}) as follows $A_{\rm RH}  \sim  (\sqrt{\Gamma_\phi M_{\rm Pl}}/ \kappa T_{\rm Max})^{-8/(4+n)}$. Since $A_{\rm RH}$ depends on $\Gamma_\phi$ and the other quantities do not depend on $\Gamma_\phi$, the above inequality involving $A_{\rm RH}$ is typically not constraining. Although we highlight here that Big Bang nucleosynthesis constraints \cite{Sarkar:1995dd} imply a limit $T_{\rm RH}\gtrsim 10$ MeV.

	\begin{figure}[t!]
		\centerline{
		\includegraphics[height=0.42\textwidth]{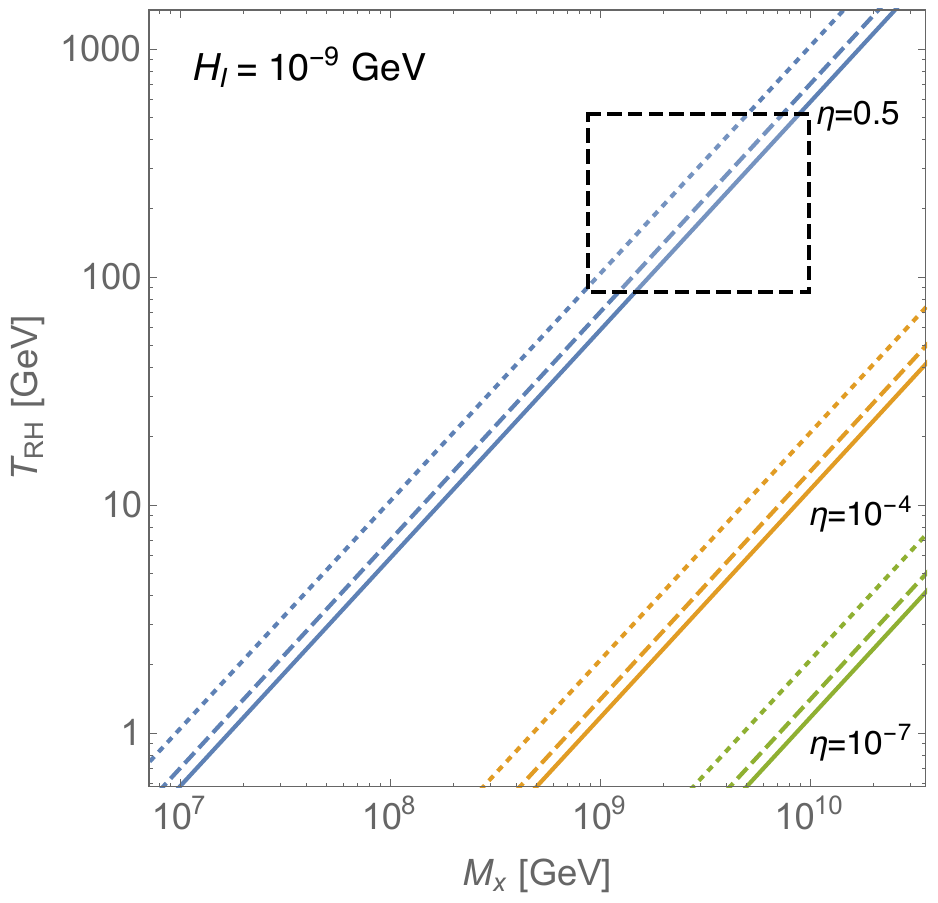}
		\hspace{3mm}
		\includegraphics[height=0.42\textwidth]{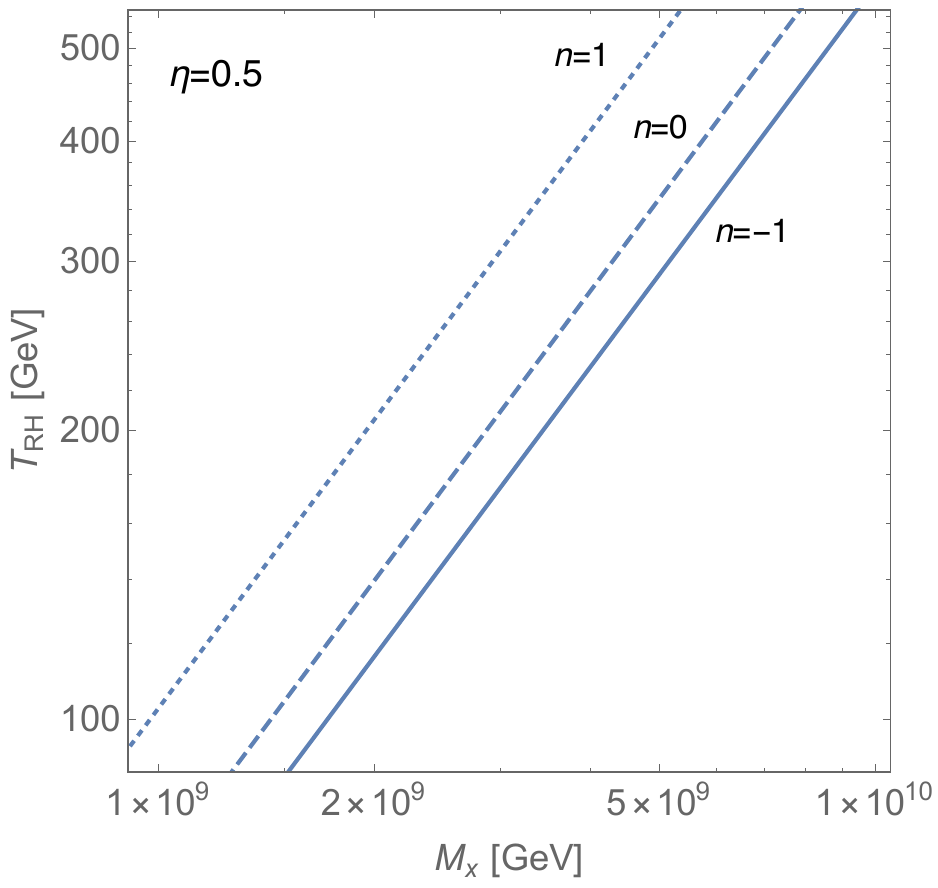}}
		\caption{Plot shows lines in the $T_{\rm RH}$-$M_X$ plane for which the dark matter relic density is reproduced via non-thermal production, following eq.~(\ref{nontrd}). Taking three  different exponents of the initial expansion rate $n=1$ (dotted), $n=0$ (dashed) and $n=-1$ (solid) and pameterising the $\phi$-dark matter branching fraction in terms of $\eta\equiv b\cdot{\rm GeV}/m_\phi$ we show three different values $\eta=0.5,~10^{-4},~10^{-7}$. The plot fixes the initial Hubble rate to be $H_I=$eV. The right panel shows an enlargement of the dashed rectangle of the left panel and illustrates the difference between $n=0,1$, and $-1$. \label{fig:4} \vspace{-3mm}}
	\end{figure}

In Figure \ref{fig:4}  we illustrate some example parameter ranges which reproduce the observed dark matter relic density for the case of  non-thermal production without chemical equilibrium for $n=0,1,-1$.  In particular, we highlight that changes in $n$ have a modest impact on the appropriate $T_{\rm RH}$ required to reproduce the relic density, however there is a great degree of freedom in  $\eta\equiv b/m_\phi$ which can lead to substantially larger impacts on the required value of $T_{\rm RH}$ needed to reproduce the observed dark matter relic density.

Note that if the branching fraction of $\phi$ decays (controlled by $b$) is sufficiently large, the dark matter will enter equilibrium in which case the contribution from non-thermal production is reduced due to dark matter annihilations. The production of dark matter due to $\phi$ decays can maintain the dark matter at an equilibrium abundance past $T\sim M_X$ and dark matter only freezes out at $T\sim T_{\rm RH}$, when dark matter production ceases. Thus the period of freeze-out occurs during the era of radiation domination. As argued in \cite{Gelmini:2006pw} (Case 4), this leads to a scaling of the radiation dominated relic density due to the difference in the entropy density between $T_{\rm RH}$ and the radiation domination freeze-out temperature $T_{\rm FO}$. Specifically, the dark matter relic abundance will be $\Omega_X\sim(T_{\rm FO}/T_{\rm RH})\Omega_{\rm RD}$ where $\Omega_{\rm RD}$ is the dark matter abundance expected from thermal freeze-out during radiation domination. Since in this case freeze-out occurs during radiation domination, the relic abundance will be largely insensitive to the temperature dependance of the initial expansion rate.


\section{Concluding Remarks}
\label{sec4}

A myriad of scenarios exists in which the early universe is not immediately radiation dominated but goes through periods with expansion rates different to the commonly assumed relationship $H\propto T^2$. In this work we have focused on a previously unstudied case in which the early universe is dominated by some state $\phi$ which leads to a general expansion rate of the form $H\sim T^{2+n/2}$, but due to the fact that  $\phi$ is decaying there is a subsequent transition to $H\sim T^4$. Notably, the form of the initial expansion rate leaves a lasting imprint on relic densities established while $H\propto T^4$, because the value of the exponent $n$ changes the temperature evolution of the Standard Model thermal bath

The prospect of the dark matter relic abundance being established during a period of entropy injection following an early matter dominated era was originally studied in influential papers  of  Giudice, Kolb \& Riotto \cite{Giudice:2000ex} and Gelmini \& Gondolo \cite{Gelmini:2006pw}. This was later  adapted to the case of a period of decays following kination domination by Visinelli \cite{Visinelli:2017qga}. In this work we have further generalised to the case in which the initial epoch has a general expansion rate  of the form $H\sim T^{2+n/2}$. We have highlighted how the choice of $n$ propagates into the cosmology of the era of significant entropy injection from $\phi$ decays and into calculations of the dark matter relic density. While freeze-out during reheating (\S\ref{3.2}) is largely insensitive to the initial expansion rate, the abundances of dark matter produced via freeze-in (\S\ref{3.1}) or non-thermal production without equilibrium (\S\ref{3.1}) are sensitive to the value of the exponent~$n$. 

Notably, the temperature dependance of the initial expansion rate can significantly impact the form of the  dark matter relic density. Since such variant cosmologies can alter the predicted dark matter relic density  they have previously been used to adjust the freeze-out abundance or evade experimental constraints, see e.g.~\cite{Gelmini:2006pw,Salati:2002md,Profumo:2003hq,Pallis:2005hm,Bramante:2017obj,Randall:2015xza,Bernal:2018ins,Gelmini:2006pq,Gelmini:2006mr,Arbey:2009gt,Hardy:2018bph,    Roszkowski:2014lga,Roszkowski:2015psa,DiMarco:2018bnw,Carr:2017edp,Bernal:2018kcw,Bhattacharyya:2018evo,Bernal:2018qlk,M.Yu.Khlopov1,M.Yu.Khlopov2,Pallis:2004yy,Drees:2017iod,Drees:2018dsj}. It would be interesting to examine how these specific particle physics models (such as the bino, neutralino, and Higgs portal) vary in the context of the generalise scenario outlined here. Moreover, in future work we plan to explore to what extent the temperature dependance of the early expansion rate imprints on cosmological parameters and observables, such as deviations in the matter power spectrum \cite{Fan:2014zua,Erickcek:2011us,Redmond:2018xty}, and whether there is a degeneracy between the initial temperature dependence of $H$ (the value of $n$) and the magnitude of the initial expansion rate $H_I$. Also, while the results presented here assume that thermalisation occurs quickly, it would be interesting to consider the converse case, and to track the potential impact of thermalisation on the dark matter abundance, generalising the work of \cite{Harigaya:2014waa,Mukaida:2015ria,Harigaya:2019tzu} to different initial expansion rates.

\vspace{3mm} {\bf Acknowledgments.}
CM is funded by FONDECYT (project 1161150), CONICYT-PCHA/Doctorado
Nacional/2018-21180309 and thanks the University of Illinois at Chicago for hospitality. JU is grateful to New College, Oxford and the Simons Center for Geometry and Physics (Program: Geometry \& Physics of Hitchin Systems) for hospitality and support.


\end{document}